\documentclass[11pt,twoside]{article}
\usepackage{newpasp,epsf}
\def\edcomment#1{\iffalse\marginpar{\raggedright\sl#1\/}\else\relax\fi}
\reversemarginpar

\begin{document}
\title{Confronting Hydrodynamic Simulations Of Relativistic Jets With Data:
What Do We Learn About Particles \& Fields?}
\author{Philip A. Hughes}
\affil{Astronomy Department, University of Michigan, Ann Arbor, MI 48109-1090}

\begin{abstract}
We review recent relativistic hydrodynamic simulations of jets, and their
interpretation in terms of the results from linear stability analysis.  These
studies show that, interpreted naively, the distribution of synchrotron
intensity will in general be a poor guide to the physical state (density and
pressure) of the underlying flow, and that even if the physical state can be
inferred, it, in turn, may prove to be a poor guide to the source dynamics,
in terms of the transport of energy and momentum from the central engine.
However, we demonstrate that an interplay of simulation and linear stability
analysis provides a powerful tool for elucidating the nature and character of
structures that jets may sustain. From such studies we can explain the
complex behavior of observed jets, which manifest both stationary and
propagating structures, without recourse to ad hoc macroscopic disturbances.
This provides a framework for the interpretation of multi-epoch total
intensity data wherein an understanding of the character of individual flow
features will allow the effects of physical state and dynamics to be
deconvolved.
\end{abstract}

\section{Introduction}
While computational fluid dynamic simulations of astrophysical jets have been
performed for two decades, it is only since the mid-1990s that this approach
has been applied to non-steady albeit non-MHD {\it relativistic} flows (van
Putten 1993; Duncan \& Hughes 1994; Mart\'\i, M\"uller, \& Ib\'a\~nez 1994,
Mart\'\i\ et al., 1995, 1997; Komissarov \& Falle 1998; Rosen et al. 1999).
Such studies have a dual significance. Firstly, that the relativistic Euler
equations are inherently more nonlinear than their classical cousins, that
the Lorentz factor couples the flow along each coordinate direction, and that
there is no intrinsic limit to either shock thinness or compression, imply
that the distributions of density, pressure and velocity will be quite
different from those computed classically. Secondly, such computations
provide a physically meaningful picture of the flow dynamics, from which
Doppler shift, boost, aberration and time delays may be inferred, allowing
the production of simulated source maps that incorporate all the pertinent
relativistic effects.

With the parallel advances in the area of VLBI/P mapping, which have led to
data sets with hitherto unseen spatial, temporal and spectral resolution, and
dynamic range, it might be supposed that, by confronting these new
simulations with the current and emerging data, the state and dynamics of
extragalactic jets on a range of length scales could now be probed in
detail.  The goal of this review is to urge caution on those involved in this
enterprise, a warning to the curious (James 1984) on the periphery of the
subject, and encouragement that progress can be made, but that such progress
is probably contingent on our acquiring a deeper understanding of jet
dynamics -- through, for example, carefully integrating analytic studies of
jet stability with studies based on CFD simulations.

A good case can be made that, at least on observable scales, the magnetic
field does not play a dynamically important role; strong evidence exists for
a turbulent field in the quiescent state of both BL~Lacs and QSOs (Jones et
al. 1985; Hughes, Aller \& Aller 1989a,b).  The magnetic field responds
readily to the underlying flow, and VLB polarimetry will surely provide an
invaluable tool for probing jet curvature, shear and the structure of
internal, in general oblique, shocks, through the imprint of these features
on the field, and thus on the polarized emission.  However, given the field's
likely insignificance as regards the flow dynamics, and the fact that a
rigorous calculation of the flow emissivity would require a knowledge of the
evolution of the distribution function of the radiating particles as much as
a knowledge of the magnetic field (Tregillis et al. 1999; see also Tregillis,
Jones \& Ryu these proceedings), for the present discussion we shall consider
purely hydrodynamic flows, and the total intensity only.

\section{The Intensity Distribution}
\begin{figure}
\plotone{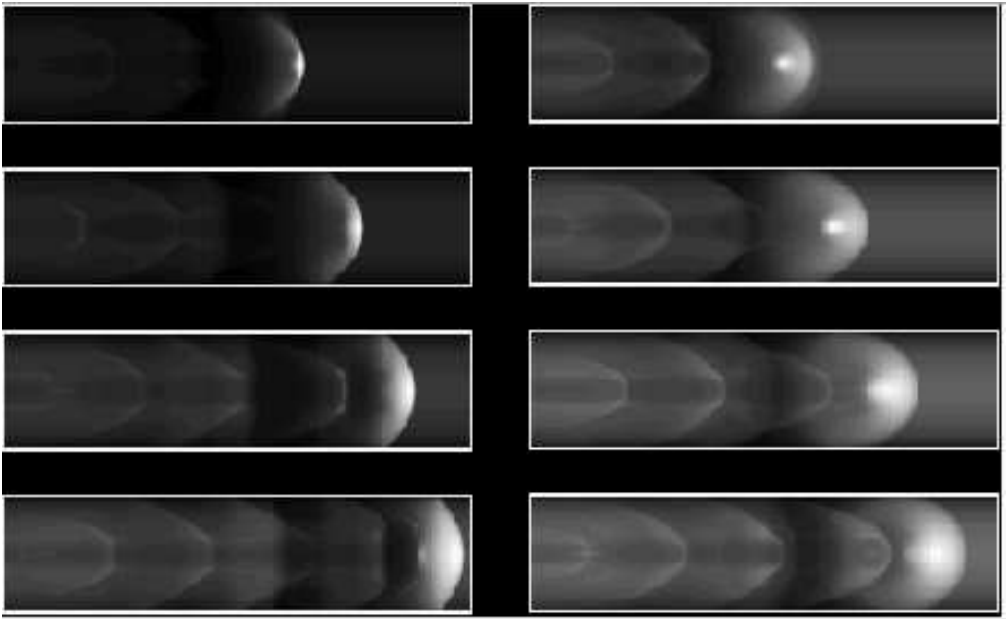}
\caption{Simulated intensity maps for the late time evolution of a flow with
$\gamma=10$, seen at an angle $90^{\circ}$ to the flow direction. Left:
without retarded time effects; Right: with retarted time effects included.}
\end{figure}
\begin{figure}
\plotone{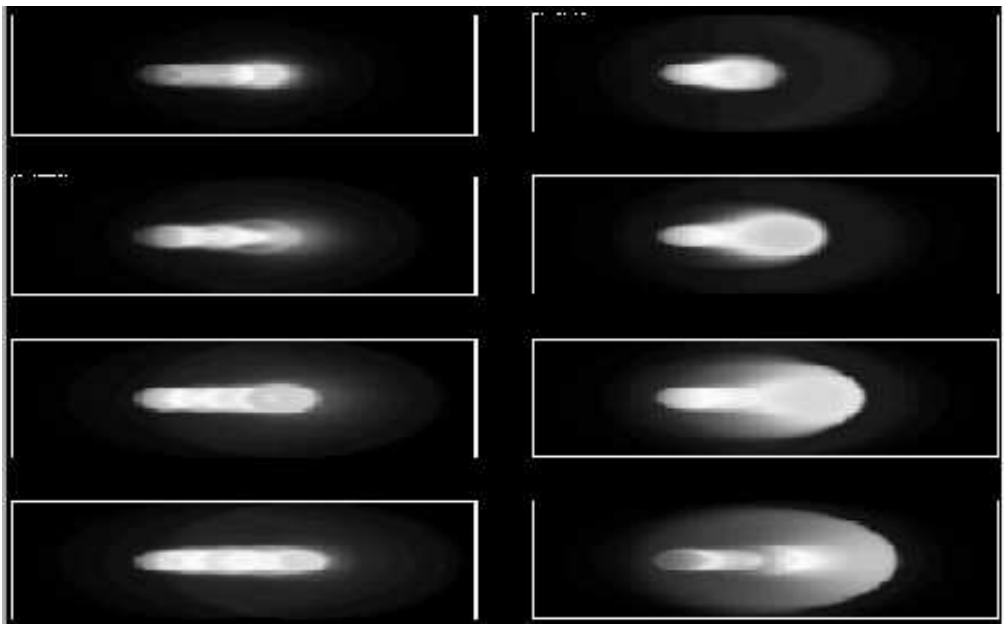}
\caption{Simulated intensity maps for the late time evolution of a flow with
$\gamma=10$, seen at an angle $10^{\circ}$ to the flow direction. Left:
without retarded time effects; Right: with retarted time effects included.}
\end{figure}
Figures~1 and 2 shows simulated maps that extend the work of Mioduszewski,
Hughes \& Duncan (1997), for an axisymmetric flow with $\gamma_{\rm max}=10$,
$\Gamma=4/3$, periodically ramped down to $\gamma\sim 1$.  The `mapping'
assumes that the radiating particle spectrum follows
$N\left(\epsilon\right)d\epsilon =N_0\epsilon^{-p}d\epsilon$,
$\epsilon_1<\epsilon< \epsilon_2$, $2<p<3$, with $N_0$ determined from
$n=\int N\left(\epsilon\right)d\epsilon$ and $e=\int
N\left(\epsilon\right)\epsilon d\epsilon$ and assumes that $e_B\sim e$.  The
intensity distribution is displayed for viewing angles of $90$ and
$10^{\circ}$ with and without retarded time employed.  The density and
pressure distributions that arise from the evolution of the inflow manifest a
series of nested bow shocks, similar in form to the pattern of intensity
displayed by the flow when seen in the plane of the sky.  However, the
intensity distribution bears little relation to the underlying
pressure/density distribution, at least for rapidly evolving, partially
opaque flows, when seen within tens of degrees of the flow direction.  The
change in source appearance with viewing angle and with the inclusion of
retarded time effects, make clear the role played by Doppler boosting and
source evolution respectively, and evidently these effects dominate over
intrinsic emissivity in determining the (evolving) intensity distribution.
The point-by-point intensity is thus a poor guide to the emissivity
intergrated along a line of sight, and in the absence of a detailed knowledge
of the flow dynamics, can be used neither to derive rest frame quantities
such as density and pressure, nor to assess the relative values of these
quantities from point to point in the source.

\begin{figure}
\plotone{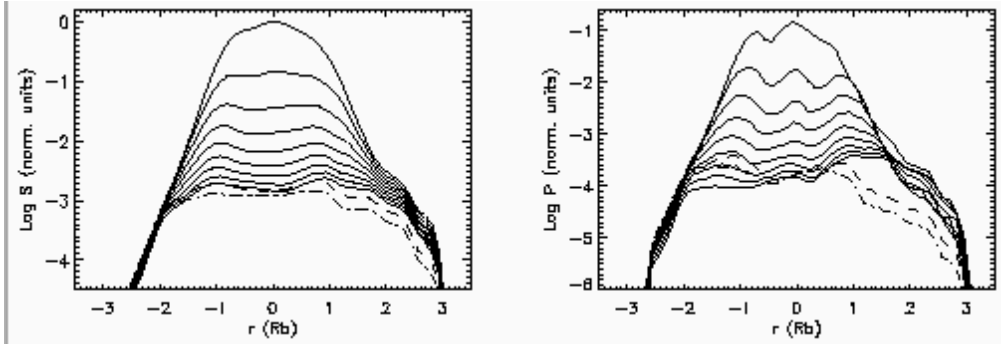}
\caption{Left: total intensity and Right: polarized intensity profiles across
a stratified jet for various viewing angles at $10^{\circ}$ intervals between
$10$ and $90^{\circ}$ (solid lines) and $-130$ and $-170^{\circ}$ (dashed lines).
Preprinted from Aloy et al. (2000) by kind permission of the author and ApJ.}
\end{figure}
Indeed, the morphology of a source may exhibit unusual characteristics,
suggestive of an anomalous field and/or particle distribution, or flow
pattern, but which in reality is merely a consequence of the interplay of a
particular pattern of intrinsic emissivity, magnetic field geometry and
Doppler boosting.  Figure~3 reproduces a figure from Aloy et al. (2000) who
have performed 3-D relativistic CFD jet simulations that produce a
significantly stratified jet structure: a fast spine surrounded by a slower,
high internal energy shear layer.  The computation is purely hydrodynamic,
but by assuming a distribution of magnetic field energy density and geometry
related to the computed flow, the authors demonstrate that at least for a
steady flow (no retarded time effects included) a significant cross-sectional
asymmetry results from the ordered component of field in the sheath, which is
a function of viewing angle, and which thus differs between jet and counter
jet.  This may well explain the curiously asymmetric structure of 1055$+$018
(Attridge, Roberts, \& Wardle 1999), and of the BL~Lac objects 0745$+$241,
0820$+$225, and 1418$+$546 (Pushkarev \& Gabuzda, these proceedings; Gabuzda,
Pushkarev \& Garnich 2000).  This is a valuable insight as it provides a
potential method for exploring the subtleties of the magnetic field
configuration in jets.  However, it also drives home the point that curious
and complex source morphologies do not necessarily have a parallel in either
the underlying particle and field distribution or in the source dynamics --
which for the example cited here constitutes a simple sheared flow and
related magnetic field.

\section{The Particle and Field Distribution}
Let us put aside for the moment the issue of how to derive the underlying
particle and field distribution from the pattern of intensity, and suppose
that such information can be extracted from multi-epoch maps.  To what extent
can we then infer the key facets of the source dynamics -- by which we mean
the way energy and momentum are carried from the central engine, and the
extent to which dissipation arising from naturally occurring instabilities and
externally imposed perturbations influence this flux of energy and momentum?

\begin{figure}
\plotone{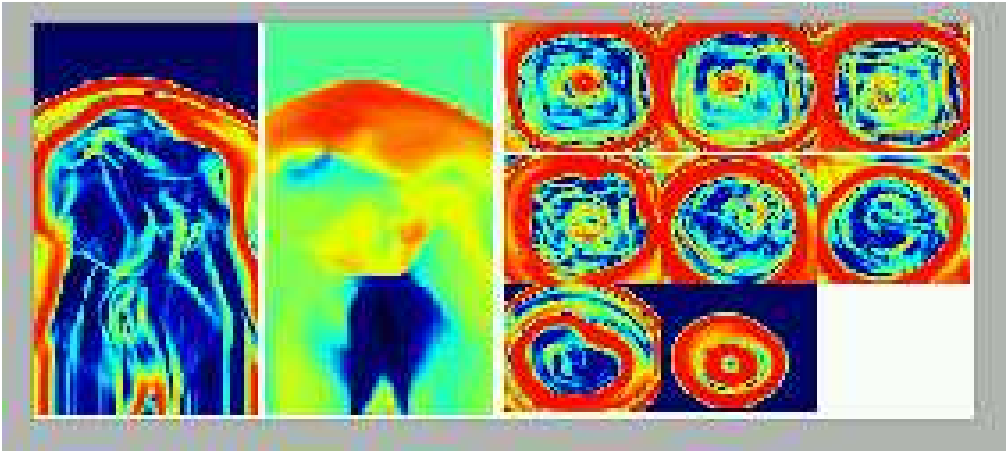}
\caption{Left: a schlieren render of the laboratory frame density in a plane
orthogonal to the inflow plane; Middle: a render of the pressure distribution
on an exponential map for the same cut as shown on the left; Right: schlieren
renders of the laboratory frame density for a series of cuts parallel to the
inflow plane at the last computed epoch of a precessed, $\gamma=2.5$ jet.}
\end{figure}
Figure~4 shows a 3-D relativistic CFD simulation performed by the author in
collaboration with M. A. Miller (Washington University) and G. C. Duncan
(Bowling Green State University) for a $\gamma=2.5$, $\Gamma=5/3$ jet
precessing on a cone of semi-angle $11.25^{\circ}$ with a frequency $0.2885$
rad measured in time units set by the inflow radius and speed.  A bubble of
high pressure gas spans both the jet and the cocoon and this bubble and the
network of shocks visible throughout the computational volume suggest that a
complex intensity pattern would result.  However, the momentum flux, ${\cal
F}=\gamma^2\left(e+p\right) v_z^2+p,$ persists to the bow, which is advancing
at a speed  $\sim 87$\% that of the unprecessed case.  Evidently, complex
maps may obscure `simple' dynamics.

\begin{figure}
\plotone{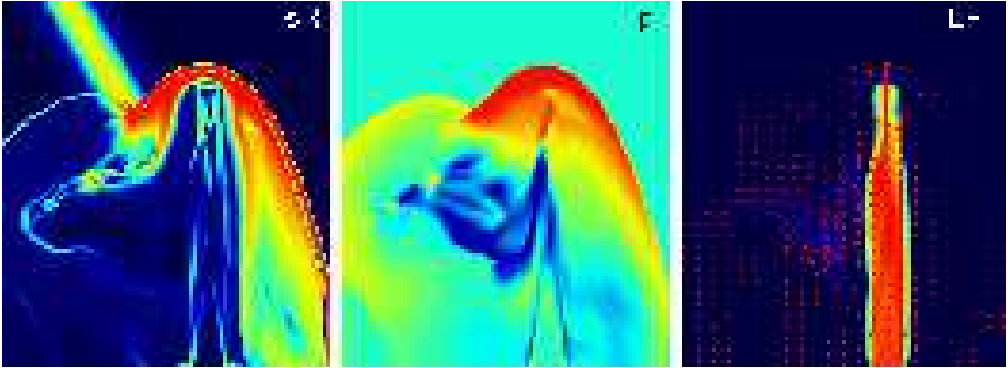}
\caption{Left: a schlieren render of the laboratory frame density in a plane
orthogonal to the inflow plane; Middle: a render of the pressure distribution
on an exponential map for the same cut as shown on the left; Right: a map of
the Lorentz factor, with velocity 3-vectors superposed, at the last computed
epoch for a $\gamma=2.5$ jet interacting with an ambient density gradient.}
\end{figure}
Figure~5 shows the results of another simulation performed in collaboration
with Miller and Duncan, in which a $\gamma=2.5$ jet impacts an ambient
density gradient inclined at $65^{\circ}$ to the flow axis, with $n_{\rm
max}/n_{\rm min}=10$.  The bow is slowed significantly by the interaction,
and the `back flow' is highly asymmetric, leading to an orthogonal {\bf
relativistic} flow with low density and pressure.  As the vectors show in the
rightmost panel, this backflow impacts the base of the jet, and is largely
responsible for oblique internal structures set up within the jet.  However,
the intensity will track the Doppler boosted jet and the high pressure bow
shock, not the `relativistic deodorant spray' feature.  Evidently, simple
maps may obscure intriguing dynamics.

\section{Understanding The Structure Of Relativistic Jets}
If the intensity distribution cannot be related in a simple way to the
underlying particle and field distributions, and given that (even if some
approach enabled these distributions to be elucidated) they may give a very
misleading impression concerning the essential flow dynamics, is it possible
to make any progress in the interpretation of total intensity maps; in
particular, by confronting the results of CFD simulations with such data?
The problems have arisen because of our attempt to relate synchrotron
intensity to physical state, and physical state to dynamics.  However, the
total intensity map contains too little information to allow this to be done
uniquely.  Further constraints are needed, and we suggest that a promising
way forward is to interpret the data in the context of an understanding of
what structures jets could and do support.  Specifically, as we shall now
show, by combining the results of CFD simulations with the results of first
order stability analysis, we can both understand the nature and origin of
features evident in the simulations, and demonstrate that, contrary to
expectation, we may use a linear stability analysis to predict the large
amplitude flow features evident in data and simulations. While simulation
provides a vital check on the predictions of linear stability analysis, and
is essential for reproducing the detailed flow pattern, the stability
analysis provides a tool for rapidly exploring a wide range of parameter
space, and simple, idealized models of flow structures.  Future application
of these studies as a tool to aid in the interpretation of data impose two
requirements. Firstly, that our ability to recognize and understand flow
features as (possibly Doppler boosted and aberrated) distinct modes of the
Kelvin-Helmholtz instability (which may have steepened to form shocks) be
tested in the simplest possible case -- slowly evolving, transparent
kiloparsec-scale flows. Secondly, that the signature of such structures in
rapidly evolving, partially opaque flows be computed for comparison with data
on the more challenging parsec-scale flows.

\section{The Structure of Axisymmetric Jets}
\begin{figure}
\plotone{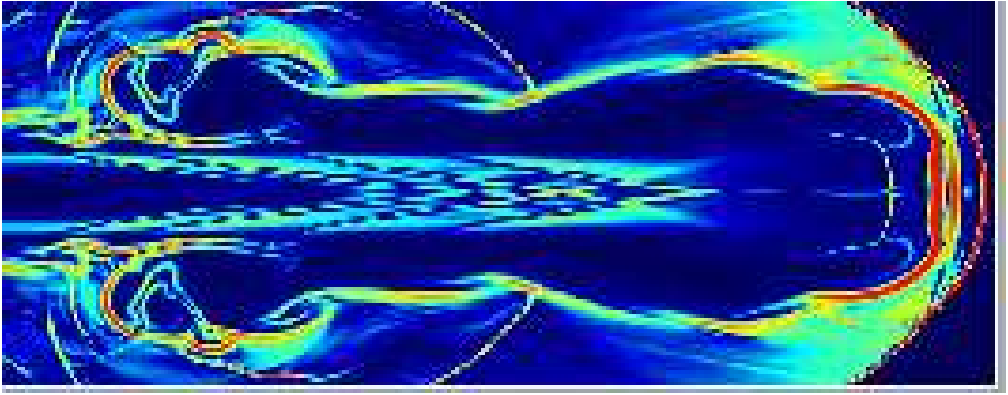}
\caption{A schlieren render of the laboratory frame density for an axially
symmetric jet with inflow Lorentz factor $\gamma=2.5$.}
\end{figure}
\begin{figure}
\plotone{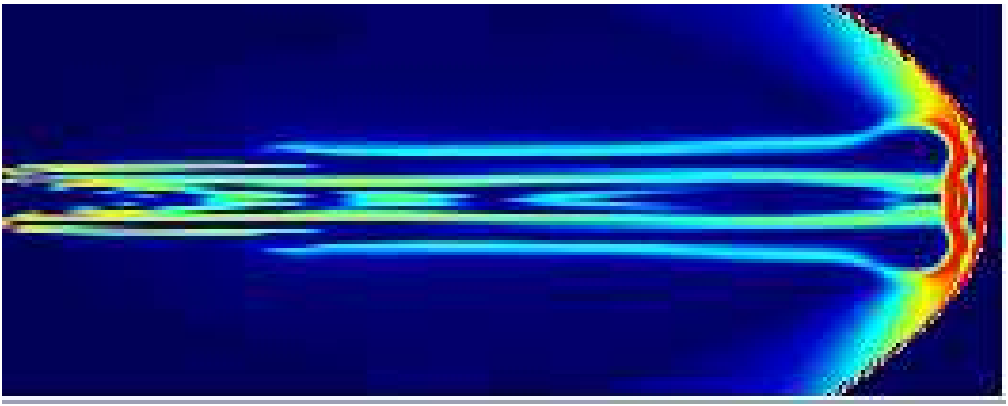}
\caption{A schlieren render of the laboratory frame density for an axially
symmetric jet with inflow Lorentz factor $\gamma=10.0$.}
\end{figure}
Figures~6 and 7 shows simulations of axisymmetric jets with $\gamma=2.5$,
$\Gamma=5/3$ and $\gamma=10.0$, $\Gamma=4/3$ respectively, performed by the
author in collaboration with Duncan. In collaboration with P. E. Hardee and
A. Rosen (University of Alabama) a first order Kelvin-Helmholtz stability
analysis (Hardee et al. 1998) has been used to identify the internal jet
structure seen in Figure~6 as due to driving of the third body mode ($B_3$)
by pressure perturbations associated with the cocoon vortices that arise due
to instability of the contact surface between shocked jet and shocked ambient
material. In contrast, the slight internal jet structure seen in Figure~7 has
been associated with driving of the first body mode ($B_1$) by the conical
pressure wave at the inlet; furthermore, the wavelength and obliquity of the
perturbation do not well match those of the excited mode, and this weak
coupling explains why the mode is barely evident.

\section{The Structure of 3-D Jets}
\begin{figure}
\plotfiddle{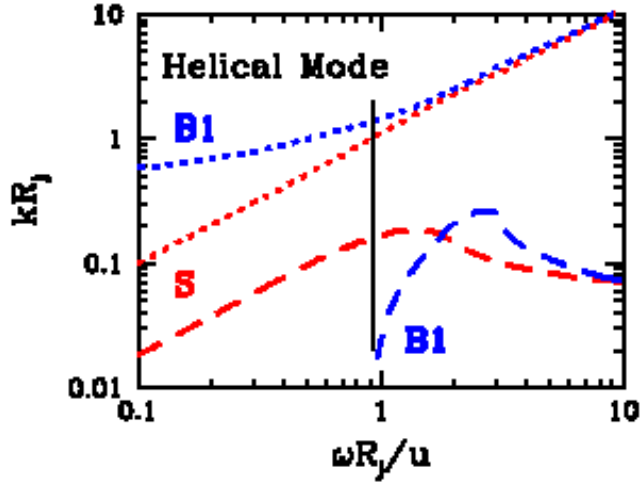}{6.0cm}{0.0}{99.0}{99.0}{-180.0}{-50.0}
\caption{The dispersion relation used to motivate the 3-D jet perturbations.
cf. Hardee 2000.}
\end{figure}
\begin{figure}
\plotfiddle{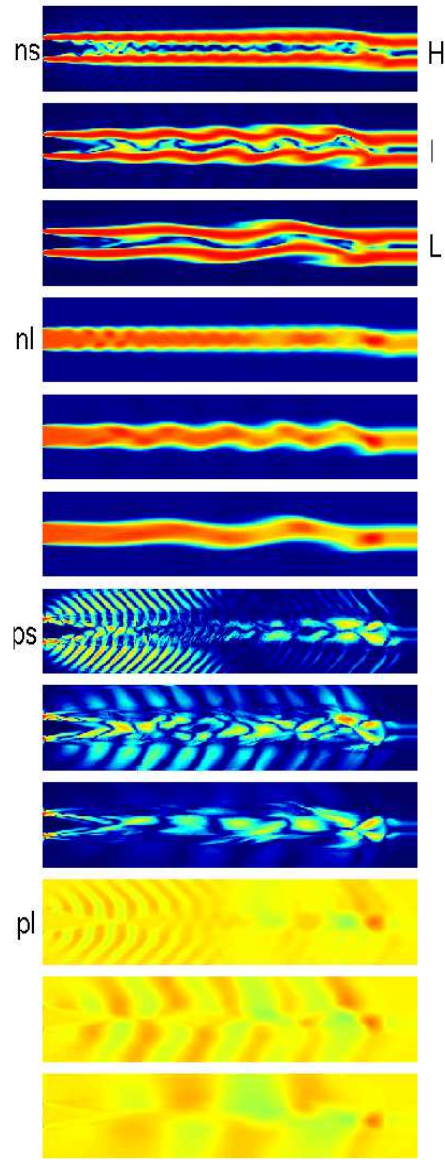}{13.0cm}{0.0}{130.0}{130.0}{-140.0}{-50.0}
\caption{Renders of rest frame density ($n$) and pressure ($p$) in schlieren
($s$) and linear ($l$) maps, for the high, intermediate and low (H, I, L)
frequency perturbations of a preexisting $\gamma=2.5$, $\Gamma=5/3$ jet.}
\end{figure}
Our success in understanding the structure of axisymmetric flows in 2-D has
prompted us to apply the same principles to 3-D jets (see Figure~8), where
instability is likely to play a more important role, as the higher order
modes -- suppressed in axisymmetry -- can manifest themselves. Figure~9 shows
a simulation performed by the author in which a pre-existing jet with
$\gamma=2.5$, $\Gamma=5/3$ has been subject to a low amplitude precessional
perturbation ($v_{\perp}<1\%v_{\rm jet}$) at the inflow. This configuration
was adopted so as to avoid having to devote valuable computational resources
to facets of the dynamics -- the contact surface and bow shock -- not
directly related to the issue of internal jet structure, and to provide an
environment in which the jet could be excited with carefully selected
perturbations. This scenario corresponds to an epoch long after the jet head
and bow shock have passed by, and the jet is cocooned by a {\it light} medium
-- the shocked jet material.  We have adopted a density ratio of $n_{\rm
jet}/n_{\rm amb}=10.0.$ Three perturbation frequencies have been explored:
low (L), with $\omega R_{\rm jet}/v_{\rm jet}=0.40$; intermediate (I) with
$\omega R_{\rm jet}/v_{\rm jet}=0.93$; and high (H), with $\omega R_{\rm
jet}/v_{\rm jet}=2.69$. The values were chosen to excite either or both the
helical surface and first body modes, using dispersion relations computed by
Hardee -- see Figure~8.

An important, but sometimes overlooked, aspect of the stability analysis is
that we can go beyond simply computing dispersion relations, and for one or a
set of modes `reconstruct' the flow pattern that arises as a consequence of
their development. Specifically, we can relate all components of the fluid
displacements at some $r$ to radial displacement at $R_{\rm jet}$, the
variations in state variables such as $p$  at any $r$ to the displacements in
fluid at that $r$, and the velocity components of perturbations at some $r$
to the time derivative of displacements in the fluid at that $r$. Thus for
some adopted surface pressure fluctuation, surface displacement, or whatever,
we can construct a detailed picture of a perturbed jet. An apparent (and
possibly severe) limitation of this reconstruction is the very nature of the
stability analysis: it assumes infinitesimal perturbations. One might thus
suspect that it would be incapable of predicting the large amplitude
structures that will result in CFD simulations. 

\begin{figure}
\plotone{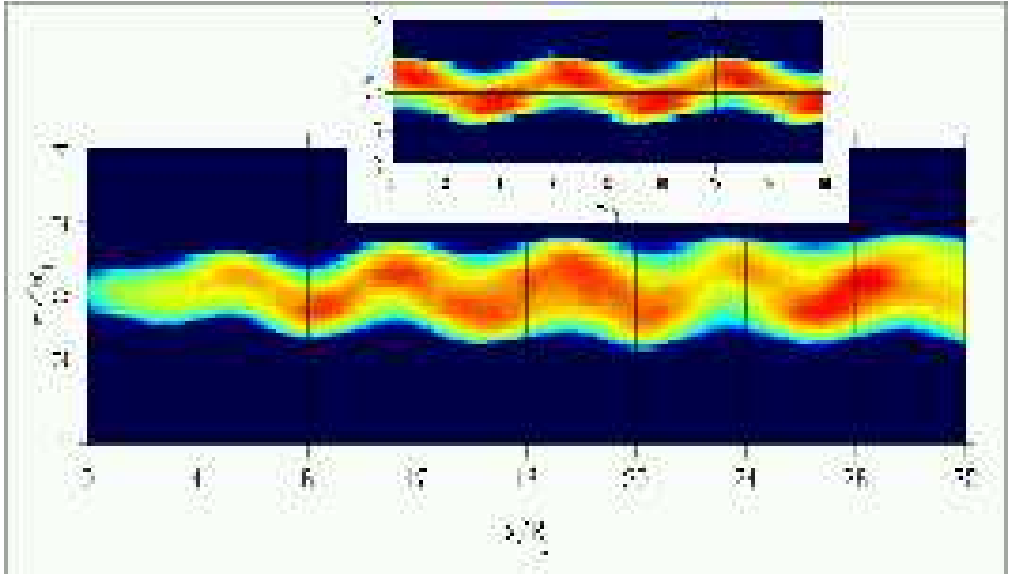}
\caption{Main: the line of sight integral of $p^2$ for the intermediate
frequency perturbation shown in Figure~9; Inset: the same diagnostic,
computed for a jet reconstructed from a first order stability analysis,
assuming that the jet structure is dominated by the helical surface mode.}
\end{figure}
To explore this issue, let us focus on case `I' described above. Figure~10
shows a comparison of the simulated flow with a reconstruction that assumes
the helical surface mode dominates, using the line of sight integral of $p^2$
as a diagnostic.  Remarkably, we see that the numerical simulation has
excited a well-defined mode of the Kelvin-Helmholtz instability, and that the
first order analysis has done an excellent job of predicting jet structure.
A full analysis of the simulations shown in Figure~9 in the context of the
normal modes analysis shows that due to excitation of the surface and first
body modes, the jet exhibits propagating structures with speeds between
$0.62c$ and $0.86c$, that fastest moving structures traveling at a speed not
much less than that of the underlying fluid flow; but furthermore, due to a
beating between these modes, a pattern of stationary structures also occurs.
Thus the complex behavior of real jets (e.g, that of M~87) may be both
reproduced and understood in terms of structures that inevitably develop in
response to small perturbations, and need not be supposed to arise from some
unspecified physics hidden in the unresolved inner regions of the source.

\section{Conclusions}
Significantly perturbed jets continue to carry energy and momentum far from
their origin, but a) the intensity maps may be a poor guide to the energy and
density distributions; b) the energy and density distributions may be a poor
guide to the source dynamics. We suggest that the interpretation of total
intensity data will be facilitated if it can be done in the context of
understanding the internal structures that jets may and do support.
Specifically, linear stability analysis aids in the interpretation of CFD
simulations, while the latter validate extrapolating stability analysis into
the nonlinear regime to predict jet structures. We thus have a powerful set
of tools with which to confront the most easily addressed data sets -- those
involving slow moving, optically thin features (e.g., as seen in M~87). The
CFD simulations suggest that jets subject to particular disturbances may
exhibit a few, identifiable modes, and if we can test our theoretical tools
on data such as that available for M~87, we will be in a position to predict
the characteristics of the modes most likely to influence jet structure, for
the more challenging case of rapidly evolving, partially absorbed flows.

\acknowledgments
This work was supported by NSF grant AST 9617032 and by the Ohio
Supercomputer Center.

\end{document}